\documentclass[a4paper,10pt]{article}

\usepackage{tabularx}
\usepackage[utf8x]{inputenc}
\usepackage{amsmath,empheq}
\usepackage{amsfonts}
\usepackage{amssymb}
\usepackage{amsthm}
\usepackage{upgreek}
\usepackage[english]{babel}
\usepackage{fontenc}
\usepackage{indentfirst}
\usepackage{subfigure}
\usepackage{authblk}
\usepackage{relsize}
\usepackage{floatrow}
\newfloatcommand{capbtabbox}{table}[][\FBwidth]
\usepackage{microtype}
\usepackage[hyphens]{url}

\usepackage[a4paper]{geometry}
\usepackage{times}
\usepackage{graphicx}
\usepackage{setspace}

\usepackage[table]{xcolor}
\definecolor{lightgray}{gray}{0.9}
\definecolor{lightb}{RGB}{228,228,228}
\usepackage{ulem}
\usepackage[backend=biber, style=numeric, url=true, doi=true, eprint=false, backref=true]{biblatex}

\usepackage{multicol}

\newcommand{\theo}{\textrm{theo}}

\title{\textbf{Non-extensive statistics in Au-Au collisions}}

\author[1]{Juliana O. Costa\thanks{E-mail: j.costa@posgrad.ufsc.br (corresponding author)}}
\author[1]{Isabelle Aguiar}
\author[1]{Jadna L. Barauna}
\author[2]{Eugenio Meg\'{\i}as}
\author[3]{Airton Deppman}
\author[1]{Tiago N. da Silva}
\author[1]{D\'{e}bora P. Menezes}

\affil[1]{\textit{\fontsize{09}{12}\selectfont Departamento de F\'{i}sica- CFM - Universidade Federal de Santa Catarina, Florian\'{o}polis - SC - CEP 88.035-972}}
\affil[2]{\textit{\fontsize{09}{12}\selectfont Departamento de F\'{\i}sica At\'omica, Molecular y Nuclear and Instituto Carlos I de F\'{\i}sica Te\'orica y Computacional, Universidad de Granada, Avenida de Fuente Nueva s/n,  18071 Granada, Spain}}
\affil[3]{\textit{\fontsize{09}{12}\selectfont Instituto de F\'{i}sica, Universidade de S\~{a}o Paulo - Rua do Mat\~{a}o Travessa R Nr.187 CEP 05508-090 Cidade Universit\'{a}ria, S\~{a}o Paulo}}

\date{ }
\bibliography{refs}

\begin{document}
\maketitle

\begin{center}
\begin{minipage}{0.8\textwidth}
\begin{abstract}
    
     Particle production yields measured in central Au-Au collision at RHIC are obtained with free Fermi and Bose gases and also with a replacement of these statistics by non-extensive statistics.
   For the latter calculation, a set of different parameters was used with values of the Tsallis parameter $q$ chosen between 1.01 and 1.25, with 1.16  generating the best agreement with experimental data, an indication that non-extensive statistics may be one of the underlying features in heavy ion-collisions.
\\
\end{abstract}
\end{minipage}
\end{center}

\begin{multicols}{2}

\section{Introduction}


In the past decades, heavy-ion collision experiments between ions, such as gold (Au) or lead (Pb), at relativistic speeds in particle accelerators \cite{Achenbach:2023pba, PHENIX:2018lia, LUO201675}, such as RHIC (Relativistic Heavy Ion Collider) and the LHC (Large Hadron Collider), have played a fundamental role in the exploration of the QCD \cite{chaudhuri2012short, LUO201675} phase diagram. 
These collisions create extreme energy conditions, providing a unique environment for the formation of the quark-gluon plasma (QGP) \cite{kalashnikov1979phase, heinz2000evidence}. This highly energetic and dense plasma recreates, for brief moments, the extreme conditions believed to have prevailed in the Universe's early stages \cite{schwarz2003first}.

However, there is no direct observation of this plasma, and its existence is inferred from the properties of the final particles that are detected in the experiments. This limitation demands the combination of experimental data and theoretical models, with thermal models playing an important role in understanding heavy-ion collisions \cite{hofmann1975report, mclerran2006td}. These models consider the QGP as a highly heated system in thermodynamic equilibrium and enable fundamental retrospective analysis to understand not only the formation of the plasma and its properties but also the generation of the resulting particles. Within this context, the use of non-extensive statistics (NES) stands out as an alternative to conventional models and statistics, offering a simple view of the particle production process. 

As the QGP formed in ultra-relativistic collisions between heavy-ions expands, it cools down. Eventually, quarks and gluons will recombine into a variety of hadrons, a process known as hadronization \cite{sweger2023recent}. As the system reaches the gaseous regime, collision times become remarkably short relative to relaxation times, resulting in chemical equilibrium.
The multiple hadron productions in these collisions offer valuable information about their nature and composition.

We present an approach to study particle production in these experiments, based on NES, contrasting with traditional approaches based on Boltzmann-Gibbs statistics. The NES statistics \cite{Tsallis:1987eu} introduces a real non-extensive parameter (\(\textbf{q}\)) that plays a crucial role in describing the degree of non-extensiveness of the system, incorporating correlation effects in complex systems. This theory offers a more complete framework for understanding emergent interactions and behaviors in these highly energetic environments.

We investigate the distributions of fermions and bosons in gold-gold (Au-Au) collision systems, using the Fermi gas model for fermions and the free Bose model for bosons. These models are particularly valuable since the free gas offers a solid basis for evaluating how particles would behave in the absence of significant interactions. NES is used to describe hadron production, using a strategy based on previous studies \cite{Menezes_2015, MEGIAS201515, Deppman_2012, Chiapparini_2019}. The application of these statistics aims to obtain an optimal fit between theoretical models and experimental data, by minimizing the value of \(\chi^2/ \text{dof}\). By considering the parameter \(\textbf{q}\), which governs the degree of non-extensivity of the system, we seek to effectively describe the interactions and transitions that lead to the formation of hadrons, enriching the understanding of physical processes involved.

This approach, based on the previous application of non-extensive statistics, represents a consolidated strategy for describing hadron formation in Au-Au heavy ion collisions \cite{Cardoso_2017, Bediaga, Deppman_2012, Cleymans_2012}. The central aim is to achieve a more accurate description of the particle dynamics under these extreme conditions by fitting theoretical models to the experimental data. 

The analyses performed in this study are based on the comprehensive experimental database from the Au-Au heavy-ion collisions performed at RHIC at $\sqrt{s}$ = 130 GeV. We have used results from STAR \cite{STAR:2001rbj, Jacobs:2001ck, Specht:2001qe, Xu2001, Huang2001}, PHENIX \cite{Ohnishi:2002rw, Messer:2002rh}, PHOBOS \cite{2003C227} and BRAHMS \cite{Videbaek2001, Bearden2001}. The main objective is to determine the optimal value of the parameter \(\textbf{q}\) that provides the most accurate description of the experimental data. To achieve this goal, three separate and complementary analyses were carried out. The first analysis focused on finding the lowest possible value of \(\chi^2\) for different variations of the \(\textbf{q}\) parameter. This meticulous process allowed an exploration of how different values of \(\textbf{q}\) affect the agreement between theoretical models based on non-extensive statistics and experimental data. In this way, the optimal value of \(\textbf{q}\) that minimizes \(\chi^2\), which represents the best fit between the model and the observed data, was identified. The second analysis focused on studying the effect of the chemical potential on the statistics of the system, considering different variations of the parameter \(\textbf{q}\).  
This approach provided valuable insights into how the chemical potential varies with the non-extensive statistics parameter.  

By studying variations in \(\textbf{q}\) it was possible to better understand how the system responds to changes in the chemical potential. Finally, the third analysis was devoted to investigating the deviations from thermodynamic equilibrium resulting from variations in the parameter \(\textbf{q}\).

The minimization shows that the optimal value of $q$ is 1.16. By comparing the results obtained with $q = 1.16$ and the conventional value of $q = 1.0$ related to the Fermi and Bose statistics, it is shown that the non-extensive statistic provides a more accurate description of the particle ratios produced in Au-Au collisions.
By exploring how the introduction of the NES statistics affects the thermodynamic equilibrium properties in heavy ion collision systems, and by taking this integrative approach, we aim at shedding a light on the underlying complexities of heavy ion collisions, bringing together the insights of non-extensive statistics and the wealth of information provided by high-energy nuclear collisions.


\section{Formalism}


In this section, we present the main equations to describe the free Fermi and Bose gases and to review the formalism underlying the use of non-extensive thermodynamics. In what follows, we will use natural units, $\hbar = c = k_B = 1$.


\subsection{Free Gases: Fermi and Bose}


The free Fermi gas is a simple model that describes particles that obey the Pauli exclusion principle and in which interactions can be neglected. 
Its Lagrangian density reads
\begin{align}
    \mathcal{L}_0^{(F)} = \bar{\psi}(i\gamma^\mu\partial_\mu - m)\psi \,,
\end{align}
where $\mathcal{L}_0^{(F)}$ is the Lagrangian density of a free fermion, the index $0$ indicates the absence of interactions. 
The terms $\bar{\psi}$ and $\psi$ represent the Dirac fields. The first term carries the kinetic part, while the second term represents the mass.

 For the purpose of the present work, particles ($\rho_{F_+}$) and antiparticles ($\rho_{F_-}$) must be treated separately \cite{PhysRevC.76.064902} and their particle densities are given respectively by
\begin{equation}
    \rho_{F_+} = (2 J + 1) \int \frac{d^3p}{(2\pi)^3} f_{F_+} \,,
    \label{denparti}
\end{equation}
and,
\begin{equation}
    \rho_{F_-} = (2J +1) \int \frac{d^3p}{(2\pi)^3} f_{F_-} \,. 
    \label{denantiparti}
\end{equation}
In equations (\ref{denparti}) and (\ref{denantiparti}),
the factor $(2J+1)$ indicates the degeneracy of the states, where $J = 1/2$ and $3/2$ represent fermions with spin$-1/2$ and spin$-3/2$ respectively. The expression $\int \frac{d^3p}{(2\pi)^3}$ integrates over all possible momenta, normalised by $(2\pi)^3$. The functions $f_{F_+}$ and $f_{F_-}$ are the Fermi-Dirac distributions for particles and antiparticles that can be expressed as 
\begin{equation}
    f_{F_\pm}= \frac{1}{e^{\beta (E\mp\mu)} + 1} \,, \label{eq:f_F}
\end{equation}
where $\pm \mu$ is the chemical potential for particles and antiparticles, $E$ is the energy, and $\beta = 1/T$ is the inverse of the temperature. 

 We now turn to the bosonic gas model, which unlike fermions, do not obey the Pauli exclusion principle.
The Lagrangian density of the free Bose gas is given by
\begin{equation}
    \mathcal{L}_0^{(B)}=(\partial^\mu \phi)(\partial_\mu \phi)- m^2\phi^2,
\end{equation}
where $\mathcal{L}_0^{(B)}$ represents the Lagrangian density of the free bosonic field, the index $0$ the absence of interactions and $\phi$ the scalar field. The first term of this Lagrangian density is the kinetic term and the second one is the mass term. The bosonic particle density is then defined below, with $J = 0$ or $1$. The particle density ($\rho_B$) is given by
\begin{equation}
\rho_B = (2J+1) \int \frac{d^3p}{(2\pi^3)} f_{B_\pm} \,.
\end{equation}
The term $\int \frac{d^3p}{(2\pi)^3}$ is the integral over the three-dimensional momentum space, normalised by the constant $(2\pi)^3$. The Bose-Einstein distribution function, $f_B$, is expressed as
\begin{equation}
f_{B_\pm} = \frac{1}{e^{\beta (E\mp\mu)} - 1} \,,
\end{equation}
which differs from the Fermi-Dirac distribution~(\ref{eq:f_F}) by the sign of the constant term in the denominator.


\subsection{Non-extensive statistics}


In the present work, we analyze the possibility to study particle collisions at high energies by using non-extensive statistics. We extend the standard exponential by \cite{Tsallis:1987eu}
\begin{equation}
    \exp{(- \beta x)} \to [1 + (q-1) \beta x]^{- \frac{q}{q-1}} \,,
\end{equation}
where $q$ is a free parameter.

NES can be used  to adjust the values of thermodynamic parameters at the chemical freeze-out, serving as an analytical tool that facilitates the description of systems outside the traditional equilibrium.

The basis of NES is the generalisation of the Boltzmann-Gibbs entropy ($S$), in such a way such it depends on a real parameter $q$ that determines the degree of non-additivity of the system, i.e.
\begin{equation}
   S_q = k \frac{1 - \sum_{i = 1} ^W p_i ^q}{q-1} \,.
\end{equation}
The non-extensive entropy for two independent systems $A$ and $B$ is given by
\begin{equation}
   S_q (A+B) = S_q(A) + S_q(B) + (1-q)S_q(A)S_q(B) \,.
\end{equation}
In this expression $q$ modulates the interaction between the entropies of the two systems. In the limit $q \to 1$, $S_q$ reduces to~$S$.

For the replacement of the Fermi and Bose statistics by the NES, the new partition function $\Xi (V,T, \mu)$ is defined as~\cite{MEGIAS201515}
\begin{equation}
    \log \Xi (V,T, \mu) = - \xi V \int _p \sum _{r = \pm } \Theta (rx) \log _q ^{(-r)} \biggl( \frac{e^{(r)}_q(x) - \xi}{e^{(r)} _q (x)} \biggr) \,,
    \label{fungranpart}
\end{equation}
where $V$ is the volume, the integral runs over the momentum, $\xi = \pm 1$ stands for bosons and fermions respectively, $x = \beta (E \mp \mu)$, $\Theta$ is the step function, and $r=\pm$ characterizes the two regimes of the $q$-exponential and $q$-logarithmic functions: $(x\ge 0)$ and $(x < 0)$. In Eq.~(\ref{fungranpart})  the $q$-logarithmic function reads
\begin{empheq}[left=\empheqlbrace]{align}
&\log_q ^{(+)} (x) = \frac{x^{q-1} -1}{q-1} \,, & \ x \geq 0 \,, 
\label{logposit}\\
&\log_q ^{(-)} (x) = \frac{x^{1-q} -1}{1-q} \,, & \ x < 0 \,,
\label{lognegat}
\end{empheq}
while the $q$-exponential function is given by
\begin{empheq}[left=\empheqlbrace]{align}
&e_q ^{(+)} (x) = [1 + (q-1)x]^{\frac{1}{q-1}} \,, & \ x \geq 0 \,,
\label{expoposit}\\
&e_q ^{(-)} (x) = [1 + (1-q)x]^{\frac{1}{1-q}} \,, & \ x < 0 \,.
\label{exponegat}
\end{empheq}
The distribution functions read~\cite{CONROY20104581}
\begin{empheq}[left=\empheqlbrace]{align}
&n_q ^{(+)} (x) = \frac{1}{\left(e_q^{(+)}(x) - \xi \right)^q} \,, & \ x \geq 0 \,, \\
&n_q ^{(-)} (x) = \frac{1}{\left(e_q ^{(-)}(x) - \xi \right)^{2-q}} \,, & \ x < 0 \,.
\end{empheq}


\section{Results}

Applying the formalism discussed in the previous sections with the purpose of describing hadron production in Au–Au collisions, the results obtained in the minimization process are then compared to experimental data.

The conservation constraints imposed on the number of baryons, the strangeness, and the electric charge are respectively~\cite{Chiapparini_2019}:
\begin{equation}
V \sum_{i}n_i B_i = Q_B = Z + N \,,
\end{equation}
with $V$ being the volume of the gas in which the particles are confined, $n_i$ is the number of particles for the $i$-hadron species, $B_i$ is the baryon number, and $Q_B$ is the baryon charge of the gas. Finally, $Z$ is the total number of protons and $N$ is the total number of neutrons in the system,

\begin{equation}
V \sum_{i} n_i S_i = Q_S = 0 \,,
\end{equation}
with $S_i$ the strangeness of the gas particles and the total strangeness $Q_S$ is zero;

\begin{equation}
V \sum_{i} n_i I_{3_i} = Q_{I_3} = \frac{Z-N}{2} \,,
\end{equation}
with $I_{3i}$ as the third component of the isospin associated with each particle, $Q_{I_3}$ as the total isospin.

The $\chi^2$ can be computed from~\cite{PhysRevC.76.064902}
\begin{equation}
    \chi^2 = \sum_i \frac{(R^{\exp}_i - R^{\theo}_i)^2}{\sigma^2_i} \,,
    \label{quiquadrado}
\end{equation}
where $R^{\exp}_i$ and $R^{\theo}_i$ are the $i$-th particle experimental and theoretical ratios, while $\sigma_i$ represents the error in the experimental data points.

\begin{figure}[h!]
    \centering
    \includegraphics[width=0.5\textwidth]{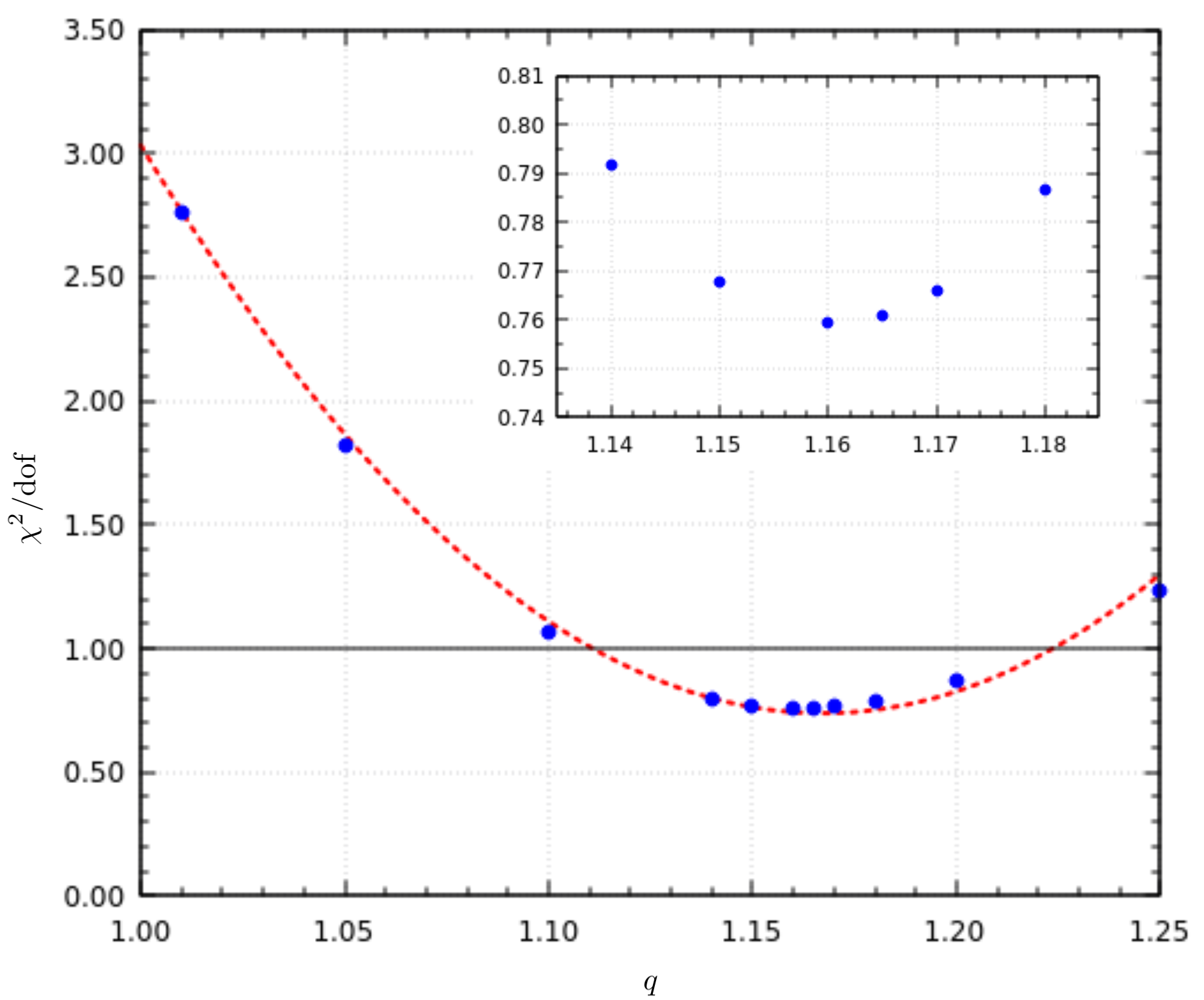}
    \caption{Results obtained from $\chi^2/\text{dof}$ as a function of $q$ in the range $1.01 \leq q \leq 1.25$, for $T=58.2$~MeV and $\mu_B=51.5$~MeV, obtained by minimization (blue). The value of $\text{dof}$ is 14, as there are 16 experimental data and 2 parameters in the fit (red). The minimum value of $\chi^2/\text{dof}$ is obtained for $ q = 1.16$. The horizontal line indicates the value of $\chi^2/\text{dof} = 1$. }
    \label{chixq}
\end{figure}

The particle ratios calculated with our model for the conventional value ($q = 1.0$) and non-extensive statistics are shown in Table~\ref{tab:1} along with experimental data obtained at RHIC from STAR, PHENIX, PHOBOS and BRAHMS Collaborations at $\sqrt{s}$ = 130 GeV \cite{Braun_Munzinger_2001} \cite{Chiapparini_2019}. 
The interval for minimizing $q$ was set between $1.01$ and $1.25$, adjusted to be close to the value of $1$ that leads to the usual statistics. This particular choice led us to the finding that $q = 1.16$ is the value that minimizes the function for $T=58.2$~MeV and $\mu_B=51.5$~MeV, as shown in Fig.~\ref{chixq}, while the range corresponding to $\chi^2/\textrm{dof} \sim 1$ is $1.11 < q< 1.22$. The particle ratios obtained with this value are also displayed in Table~\ref{tab:1}.

\end{multicols}

\begin{figure}[h!]
	\centering
	\begin{minipage}[b]{0.5\textwidth}
		\centering
		\includegraphics[width=1.0\textwidth]{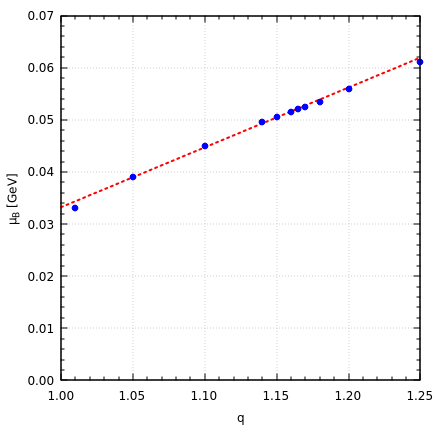}
        \label{muxq}
	\end{minipage}\hfill
	\begin{minipage}[b]{0.5\textwidth}
		\centering
        \includegraphics[width=1.0\textwidth]{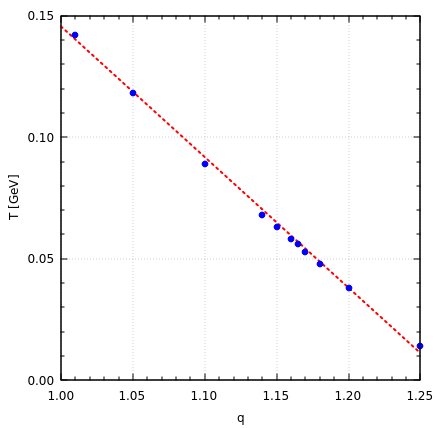}  
        \label{txq}
	\end{minipage}
	\caption{Plot of $\mu_B$ (left panel) and $T$ (right panel) as a function of $q$, both results obtained from the minimization process (blue) together with the respective fits (red). }
	\label{txq}
\end{figure}

\newpage

\begin{multicols}{2}

\color{black} The best fits corresponding to the obtained data are given by the following equations
\begin{eqnarray}
    \chi^2/\textrm{dof} &=& 3.03253 - 27.4348 (q-1) + 82.0251 (q-1)^2 \,, \nonumber \\  \\
    \mu_B &=& \left[ 33.2202 + 114.829 (q-1)  \right]\, \textrm{MeV} \,, \\
     T &=& \left[ 145.52 - 537.165 (q-1) \right] \, \textrm{MeV} \,, 
\end{eqnarray}
where $\textrm{dof} = 14$ (degrees of freedom) is obtained from the 16 experimental data and 2 fitting parameters ($T$ and $\mu_B$). 

When the Fermi and Bose statistics are used,  
the temperature $T = 148.8$ MeV and chemical potential $\mu_B = 32.5$ MeV associated with the minimum value of $\chi^2/\textrm{dof} = 2.97387$ are obtained.
For the best value of the the non-extensive statistics, $T = 58.2$ MeV, $\mu_B = 51.5$ MeV and $\chi^2/\textrm{dof} = 0.75926$ are obtained. 

\end{multicols}

\begin{table}[h!]
\centering
\begin{tabularx}{1\textwidth} { 
   >{\centering\arraybackslash}X 
   >{\centering\arraybackslash}X 
   >{\centering\arraybackslash}X 
   >{\centering\arraybackslash}X
   >{\centering\arraybackslash}X
   >{\centering\arraybackslash}X}
  \hline
    Ratio & $q = 1.01$ & $q = 1.16$ & Exp. data $\sqrt{s}$ = 130 GeV  
    & Exp. & Ref. \\ 
     \hline
     $\bar{p}/p$  & 0.648694 & 0.656767 & 0.65 $\pm$ 0.07 & STAR & \cite{STAR:2001rbj} \\
       &  &  & 0.64 ± 0.07 & PHENIX  & \cite{Messer:2002rh}\\
       &  &  & 0.60 ± 0.07 & PHOBOS & \cite{2003C227}\\
       &  &  & 0.64 ± 0.07 & BRAHMS & \cite{Bearden2001}\\
     $\bar{p}/\pi^{-}$ & 0.041363 & 0.064067 & 0.08 ± 0.01 & STAR & \cite{Jacobs:2001ck}\\
     $\pi^{-}/\pi^{+}$ & 1.008493 & 1.019848 & 1.00 ± 0.02 & PHOBOS & \cite{2003C227} \\
       &  &  & 0.95 ± 0.06 & BRAHMS & \cite{Videbaek2001} \\
     $K^{-}/K^{+}$ & 0.940661 & 0.829491 & 0.88 ± 0.05 & STAR & \cite{Specht:2001qe} \\
       &  &  & 0.78 ± 0.13 & PHENIX & \cite{Ohnishi:2002rw}  \\
       &  &  & 0.91 ± 0.09 & PHOBOS & \cite{2003C227}\\
       &  &  & 0.89 ± 0.07 & BRAHMS & \cite{Videbaek2001} \\
     $K^{-}/\pi^{-}$ & 0.235757 & 0.169266 & 0.149 ± 0.02 & STAR &\cite{Specht:2001qe} \\
     $\bar{\Lambda}^0/\Lambda^0$ & 0.689054 & 0.770251 & 0.77 ± 0.07 & STAR & \cite{Xu2001} \\
     $\bar{\Xi}^{-}/\Xi^{-}$ & 0.732019 & 0.875823 & 0.82 ± 0.08 & STAR & \cite{Huang2001}\\
     $\bar{\Omega}^{-}/\Omega^{-}$ & 0.783607 & 0.976638 &  &\\
     $\bar{\Omega}^{-}/\pi^{-}$ & 0.001389 & 0.025813 &  &\\
     $K^{0*}/h^{-}$ & 0.061003 & 0.071760 & 0.06 ± 0.017 & STAR & \cite{Xu2001} \\
     $\bar{K}^{0*}/h^{-}$ & 0.056983 & 0.062544 & 0.058 ± 0.017 & STAR &\cite{Xu2001}\\
     $h^{-}/\rho$ & 14.194047 & 7.903081 &  &\\
     $\Lambda^0/h^{-}$ & 0.013941 & 0.026713 &  &\\
     $\Omega^{-}/\Xi^{-}$ & 0.250425 & 0.871511 &  &\\
     $\Lambda^0/K^{0*}$ & 0.228535 & 0.372255 &  &\\
     $\bar{\Xi}^{-}/\Lambda^0$ & 0.222069 & 0.487735 &  &\\
     $\bar{\Xi}^{-}/\bar{\Lambda}^0$ & 0.322280 & 0.633215 &  &\\
     $\bar{\Xi}^{-}/\bar{K}^{-}$ & 0.021990 & 0.156920 &  &\\
     \hline
\end{tabularx}
\caption{Particle ratios of Au-Au collisions, considering $q$ = 1.01 and $q$ = 1.16 along with experimental data at $\sqrt{s} = 130$ GeV, are analyzed with parameters set to $T = 58.2$ MeV and $\mu_B = 51.5$ MeV.}
\label{tab:1}
\end{table}       

\newpage

As seen in Table~\ref{tab:1}, the minimum value of $\chi^2/\textrm{dof} = 0.75926$ provides a better description of particle ratios produced in Au-Au collisions than the usual statistics.

\begin{multicols}{2}

\section{Conclusions}

In this paper, we discuss the application of non-extensive statistics to describe particle production in Au-Au heavy ion collisions. This approach differs from traditional Boltzmann-Gibbs statistics by the introduction a real non-extensive parameter \(q\), which controls the degree of non-extensivity and incorporates correlation effects in complex systems.

Through the analysis of particle distributions in collision systems, we explore the effects of the introduction of the non-extensive parameter \(\textbf{q}\) on the properties of thermodynamic equilibrium, with the aim of understanding the implications of this alternative statistic in extreme conditions.

We investigate the impact of the \(\textbf{q}\) parameter on particle distributions, explore the influence of baryonic density on system statistics and analyze the variations in thermodynamic equilibrium resulting from different values of \(\textbf{q}\). By comparing the model's predictions with experimental data, we assess the suitability of the NES statistic in describing the properties observed in heavy ion collisions. Another way of corroborating this theory is by obtaining the fractions of missing particles.

As a result, although we have made progress in understanding heavy ion collisions from a NES perspective, it is clear that the field is still evolving. Future studies can focus on more refined approaches, considering a variety of additional factors that can influence particle distributions in complex collision systems. In doing so, we can continue to expand our knowledge of the properties of matter under extreme conditions and enhance our understanding of heavy ion collisions as a window into fundamental phenomena of high energy physics.



\section{Acknowledgments}

This work is a part of the project INCT-FNA Proc. No. 464898/2014-5. D.P.M. was partially supported by Conselho Nacional de Desenvolvimento Científico e Tecnológico (CNPq/Brazil) under grant 303490-2021-7. I.A. acknowledges a M.Sc. scholarship from Conselho Nacional de Desenvolvimento Científico e Tecnológico (CNPq/Brazil), J.O.C. 
acknowledges a M.Sc. scholarship from Coordenação de Aperfeiçoamento de Pessoal do Ensino Superior (Capes/Brazil)
and J.L.B. acknowledges an undergraduate research scholarship from Conselho Nacional de Desenvolvimento Científico e Tecnológico (CNPq/Brazil). The work of E.M. is supported by the project PID2020-114767GB-I00 and by the Ram\'on y Cajal Program under Grant RYC-2016-20678 funded by MCIN/AEI/10.13039/501100011033 and by ``FSE Investing in your future'', by the FEDER/Junta de Andaluc\'{\i}a-Consejer\'{\i}a de Econom\'{\i}a y Conocimiento 2014-2020 Operational Programme under Grant A-FQM-178-UGR18, by Junta de Andaluc\'{\i}a under Grant FQM-225, and  by the "Pr\'orrogas de Contratos Ram\'on y Cajal" Program of the University of Granada.

\printbibliography

\end{multicols}

\end{document}